\documentclass[twocolumn]{aastex63}

\usepackage{multirow}
\usepackage{hyperref}

\begin{document}
	
	\title{FRB 121102 Bursts at a Constant Rate per Log Time}
	\author{Elisa Tabor}
	\affiliation{Department of Physics, Stanford University,  452 Lomita Mall, Stanford, CA 94305}
	\email{etabor@stanford.edu}
	\author{Abraham Loeb}
	\affiliation{Department of Astronomy, Harvard University, 60 Garden St., Cambridge, MA 02138}
	\email{aloeb@cfa.harvard.edu}
	
	\begin{abstract}
		Despite many searches for periodicity in the repeating fast radio burst FRB 121102, the underlying pattern of bursts does not appear to be a periodic one. We report a logarithmic repetition pattern in FRB 121102 in the sense that the rate falls off inversely with time for each set of bursts. This result implies that repeating FRB sources are not necessarily associated with a pulsar, but rather could be caused by a different type of phenomenon that involves an equal amount of energy output per log time.
	\end{abstract}
	
	\keywords{radio continuum: transients}
	
	\section{Introduction} \label{sec:intro}
	Fast radio bursts (FRBs) are bright transients that last roughly a millisecond \citep{Lorimer2007, Thornton2013}. Based on their large dispersion measures and the observed redshifts of several of their host galaxies, most of the detected FRBs are believed to originate at extragalactic distances \citep{Chatterjee2017, Keane2016, Tendulkar2017}. Recently, the Canadian Hydrogen Intensity Mapping Experiment (CHIME), the Five-hundred-meter Aperture Spherical radio Telescope (FAST), and other surveys have been reporting many new bursts \citep{Fonseca2020, Li2019}.
	
	Some FRBs have been found to repeat, while others have only been detected once, and we do not have enough information to know whether the two types are different populations or whether all FRBs will eventually repeat {\citep{Caleb2019, James2019}}. In this paper, we will focus on the known repeaters in an attempt to understand some of the underlying properties of the engine that powers the bursts. 
	
	Two FRBs have been found to be modulated on particularly long periods. CHIME detected a 16 day periodicity in FRB 180916, with bursts arriving in a four day phase window, {i.e. several bursts were detected over the course of four days and none were reported the other 12 days for several periods of this FRB} \citep{CHIME2020, Marthi2020}. The other repeater, FRB 121102, is the focus of this \textit{Letter}. It is the most studied repeating FRB and was found to have a period of 157 days, with an 88 day active phase \citep{Rajwade2020}. {The period was found such that every reported observation of a burst from FRB 121102 fits within an active period and every non-detection fits within an inactive period.} However, there was not enough data collected to conclusively demonstrate that there could not be a smaller modulation period than that reported (see Figure 2 of \citealt{Rajwade2020}). Bursts from FRB 121102 originate in a star-forming region on the outskirts of a dwarf galaxy at redshift z = 0.193 \citep{Chatterjee2017, Tendulkar2017, Bassa2017, Marcote2017}.
	
	Many models have been proposed to explain the sources producing the FRBs, and the most popular one for repeating FRBs describes them as pulses from magnetars, which are neutron stars with extremely strong magnetic fields \citep{Munoz2020, Katz2020, Levin2020}. This is motivated in part by the recently discovered FRB originating in a Galactic source \citep{Scholz2020, Bochenek2020}, which demonstrates some periodicity as well \citep{Grossan2020}. However, the luminosity of this FRB was $\sim 10^3 $ times too small for it to be of the same population as the ones detected at cosmological distances \citep{Margalit2020, Beniamini2020}. {Other theoretical models that could explain periodicities involve orbital motion in binary systems \citep{CHIME2020, Rajwade2020} or the precession of neutron stars \citep{Levin2020}.}
	
	The organization of this \textit{Letter} is as follows. In section 2 we present the data set we used and our methods for data analysis. We describe the resulting fits to each data set in section 3, and finally the implications of these findings in section 4.
	
	\section{Methods}
	{Although there exist many sets of data on bursts from FRB 121102 \citep{Spitler2016, Scholz2016, Scholz2017, Michilli2018, Gajjar2018, Gourdji2019, Oostrum2020, Caleb2020, Cruces2020}, most sets have between 10 and 25 detected bursts in any single observation period, too few for any firm statistical inference.} However, the Breakthrough Listen group was able to detect 72 additional bursts to the original 21 reported by \cite{Gajjar2018} using machine learning methods, totaling 93 points in a five hour period \citep{Zhang2018}. This is the data set we will be focusing on.
	
	Any observation period must have started at a specific time, which was unlikely to be the start of the set of bursts, so we include a timescale $ t_0 $, which roughly characterizes the appearance time of the first unseen burst in the series of bursts under consideration, among our floating parameters. In analyzing the data, we used the scipy optimization {\footnotesize CURVE\_FIT} package in python\footnote{ \url{https://docs.scipy.org/doc/scipy/reference/generated/scipy.optimize.curve\_fit.html} } to develop a prediction for where the FRBs should land given each value of $ t_0 $. To find the standard deviation of these predictions, we binned the data points, evening out the Poisson fluctuations within each bin, and compared the bins against the resulting fit. We applied this algorithm to find the best fit curve and its error for each data set.
	
	\begin{figure}[h]
		\epsscale{1.1}
		\plotone{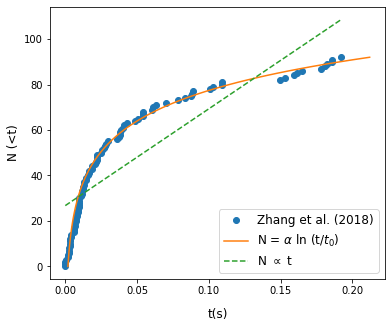}
		\caption{Cumulative number of bursts as a function of observed time in seconds. The orange solid curve represents the logarithmic fit to the data points from \cite{Zhang2018} based on equation (1), with best fit values of $\alpha=18.1$ and $t_0=121$s. The green dashed line shows an example of a periodic signal, which is unable to fit the data. The difference in their goodness of fit is illustrated by their reduced $ \chi^2 $ values, as we have $ \chi_{\nu}^2 = 7.37 $ for the orange solid curve and $ \chi_{\nu}^2 = 1040 $ for the green dashed line.}
		\centering
		\label{fig:curve}
	\end{figure}
	
	\section{Results}
	The best fit for the Breakthrough Listen data \citep{Zhang2018} is presented in Figure \ref{fig:curve}. We fit the data to the formula,
	\begin{equation}
	N(<t) = \alpha\ln(t/t_0),
	\end{equation}
	for $ t \gtrsim t_0 $, with $ N $ being the number of bursts and $ t $ the time since the start of the observation period. Our best fit values are $ \alpha=18.1 ^{+0.630}_{-0.663} $ and $ t_0=121 ^{+20.6}_{-20.3} $ s. The first in this series of 93 bursts started when $ N=1 $ at $ t=e^{1/\alpha}t_0 = 128^{+21.5}_{-21.2} $ s. Since the duration of a single burst is $ \sim $ 1 ms, this implies an initial duty cycle of $ \sim 10^{-5}$.
	
	{The orange solid curve in Figure \ref{fig:curve} represents the logarithmic fit to the data points from \cite{Zhang2018} based on equation (1), while the green dashed curve represents a periodic underlying signal. The difference in their goodness of fit is illustrated by their reduced $ \chi^2 $ values, as we have $ \chi_{\nu}^2 = 7.37 $ for the logarithmic fit and $ \chi_{\nu}^2 = 1040 $ for the periodic fit.}
	
\begin{figure}
		\epsscale{1.2}
		\plotone{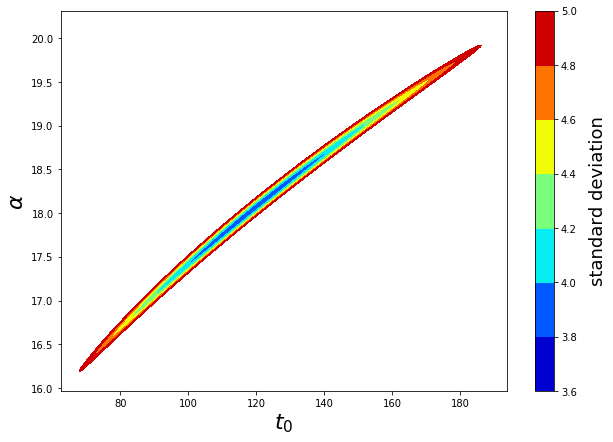}
		\caption{Contour plot demonstrating the impact of the initial timescale of the bursts ($t_0$) on the value of $\alpha$ in equation (1), where the colors represent the standard deviation, applied to the data points from \cite{Zhang2018}.}
		\centering
		\label{fig:colorplot}
	\end{figure}

	Since
	\begin{equation}
	N(<t) = \alpha\ln(t/t_0) = \alpha\ln(t) - \alpha\ln(t_0),
	\end{equation}
	taking the derivative with respect to $ \ln(t) $ yields
	\begin{equation}
	\alpha = \frac{dN}{d\ln(t)} = t\frac{dN}{dt},
	\end{equation}
	so our final result for the burst rate of FRBs is
	\begin{equation}
	\frac{dN}{dt} = \frac{\alpha}{t}.
	\end{equation}
	Figure \ref{fig:colorplot}  shows the confidence contours in the ($ \alpha,\ t_0$) plane, where the colors represent the standard deviation. The plot demonstrates how closely correlated $ \alpha $ and $ t_0 $ are.
	
	We also ran a simulation to check whether we could fit the rest of the existing data (from \citealt{Zhang2018}, \citealt{Rajwade2020}, and \citealt{Cruces2020}) to the pattern shown in our results. In our simulations, we assumed the theoretical data would appear in sets of logarithmic curves, and that the starts of these sets would either be constant (separated by 0.2 to 10 days) or random (selected from a range of 0-$N$ days where $N$ takes on integers 1 through 5). We started from MJD 57991.41, which is the start date of the data published by \cite{Zhang2018}, and we created a model set of data points that consists of curves resembling the orange solid line in Figure \ref{fig:curve}. One caveat lies in the possibility that the $ \alpha $ and $ t_0 $ found in Figure \ref{fig:curve} may be frequency dependent, but for the sake of this analysis we adopt a single frequency independent value for them. {Under this assumption, we were able to simultaneously compare detections at different frequencies and sensitivities by computing the error between the measured MJD and the closest predicted MJD.}
	
	In the constant separation scenario, this model set of data points was repeated every fixed period of time, which we varied as a free parameter to minimize the error relative to the time-tags of the detected bursts. In the second scenario, our free parameter was the number of days the random increment could be selected from, and the starts of each set of bursts were accordingly separated by random increments from that range. When the random increment is selected from 4 or less days, the error is small enough that we should not rule out this model.
	
\begin{figure*}
		\epsscale{1.15}
		\plotone{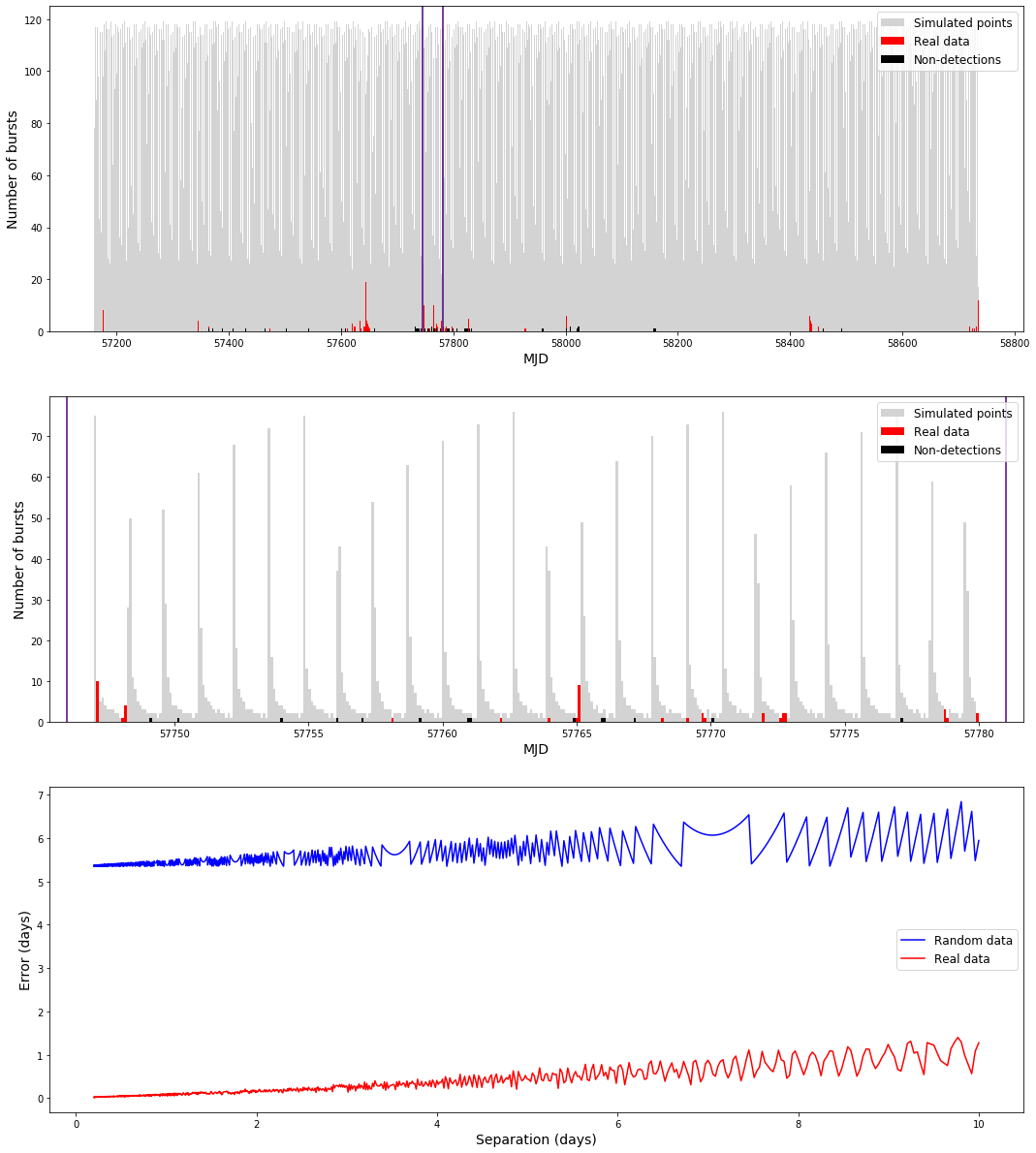}
		\caption{A simulation of a fit to the data from \cite{Zhang2018, Rajwade2020, Cruces2020}. Panels A and B are examples of a case with a steady separation of 1.3 days between sets, where the red represents real data, the black represents non-detection periods, and the gray represents the simulated set of points. The violet vertical lines demonstrate the area of Panel A that is magnified in Panel B. Panel C demonstrates the error as a function of time separation between sets of bursts for the complete activity cycle, corresponding to Panel A. The red again represents real data and the blue curve represents a random set of observations that had a uniform probability of appearing within the active periods mentioned in \cite{Rajwade2020} and \cite{Cruces2020}.}
		\centering
		\label{fig:sim}
	\end{figure*}

	A histogram of the simulation for the constant separation scenario is shown in Panels A and B of Figure \ref{fig:sim}, where the red represents real data, the black represents non-detection periods, and the gray represents the simulated set of points. The violet vertical lines demonstrate the area of Panel A that is magnified in Panel B. For each real observed data point, we found the closest simulated point and used the difference between the two time tags as our error. Our final error per point was the norm of this set of errors (the square-root of the sum of the squares divided by the number of points), and this final result is shown by the red curve in Panel C of Figure \ref{fig:sim} for various separations from 5 hours to 10 days. All separations under a day gave an error better than 0.1, so we were not able to conclusively find the best possible separation between sets of bursts. We also created a random set of observations that had a uniform probability of appearing within the active periods mentioned in \cite{Rajwade2020} and \cite{Cruces2020}, which is shown by the blue curve in Panel C of Figure \ref{fig:sim}.
	
	We also compared our modeled simulation against the non-detection periods published by \cite{Rajwade2020} and \cite{Cruces2020}, in which no bursts were detected. Whenever the model predicts a burst in such a non-detection period, the prediction is at least off by the duration of time between the closer edge of the non-detection period and the predicted burst, so we included this as additional error in our model. We have found that there is no information in these non-detection periods in terms of error optimization.
	
	\section{Discussion and Conclusions}
	An implication of these findings is that FRB 121102 is not periodic, but rather it follows a logarithmic repetition pattern, where the rate falls off as one over time for each set of bursts. This could plausibly explain the vacant, inactive regions in Figure 2 of \cite{Rajwade2020} since beyond a certain amount of time, there are so few bursts that the probability of measuring one goes to zero. However, if $ t_0 $ is on the order of hundreds of seconds, there may be many smaller sets of bursts within each period reported by \cite{Rajwade2020}. Another possibility is that FRB 121102 emits equal amounts of energy per log-time, but the emission pattern still follows the overall \cite{Rajwade2020} modulation periodicity of 157 days, thereby fitting the rest of the observations found over the past four to five years. More data and monitoring of the source is necessary to come to a more firm conclusion.
	
	Another implication is that this lack of periodicity is not consistent with the idea that repeating FRBs are a type of pulsar \citep{Beniamini2020, Munoz2020}, rather they might be caused by a different type of phenomenon that involves an equal amount of energy output per log time. This could be indicative of a process with a characteristic timescale of $ t_0 $. For example, the source of the FRB could be an object that charges up and then discharges with equal amounts of energy output per logarithmic time interval.
	
	{A potential concern with the model is that the data from \cite{Cruces2020} contains a set of 24 observations in a 7 hour period in which the rate significantly increases in the second half of the observations. Assuming the sensitivity of the data collection methods remained constant, this would imply a change in the intrinsic rate of bursting, unless within that 7 hour period one cycle of bursts ends and another begins.}
	
	A few remaining open questions are the effect of the detection threshold of the observations as there might have been many fainter bursts that were missed, possibly affecting the pattern we found, and the potential effect of frequency on the underlying burst pattern.
	
	In conclusion, we have shown that some repeating FRBs may not send periodic bursts, rather the bursts could be arriving at a logarithmic rate. Our equation (1) can be tested by upcoming data on repeating FRBs in the near future \citep{Li2019, Hashimoto2020}.
	
	A roughly constant burst rate in log time was previously noted by \cite{Wadiasingh2019}.

	\acknowledgments This work was supported in part by a grant from the Breakthrough Prize Foundation (for A.L.) and by a summer internship from Harvard's Institute for Theory and Computation (for E.T.). We thank Dunc Lorimer, Kshitij Aggarwal and Julian Munoz for insightful comments on an early draft of the paper. We also thank the anonymous referee, whose important suggestions improved the clarity of the manuscript.
	
	\newpage
	\bibliographystyle{aasjournal}
	\bibliography{FRB_121102}
	
\end{document}